\documentclass[letterpaper,12pt,oneside,onecolumn]{article}%
\usepackage{geometry}%
\usepackage{titlesec}%
\titleformat{\section}[hang]{\normalfont\normalsize\bfseries}{\thesection}{12pt}{\centering}%
\titleformat{\subsection}[display]{\normalfont\normalsize}{\thesubsection}{12pt}{\underline}%
\titleformat{\subsubsection}[runin]{\normalfont\normalsize}{\thesubsubsection}{12pt}{\underline}%
\newcommand{\PaperTitle}[1]{%
\begin{center}%
    \begin{large}%
        \textbf {#1} \\%
    \end{large}%
\end{center}%
}%
\newcommand{\AuthorList}[1]{%
\begin{center}%
    {#1} \\%
\end{center}%
}%
\newcommand{\AuthorAffiliations}[1]{%
\begin{center}%
    {#1} \\%
\end{center}%
}%
\newcommand{\Keywords}[1]{%
\begin{center}%
   Keywords: {#1} \\%
\end{center}%
}%
\usepackage{epsfig}
\begin{document}%
\PaperTitle{Interatomic forces for transition metals including magnetism}%
%
%
\AuthorList{G.J.Ackland$^1$, D.J.Hepburn$^1$ and J.Wallenius$^2$}%
\AuthorAffiliations{$^1$ SUPA, School of Physics, The University of Edinburgh, Mayfield Road, Edinburgh EH9 3JZ, UK\\
$^2$Reactor Physics, Royal Institute of Technology, 106 91 Stockholm, Sweden}%
\section{Abstract}We present a formalism for extending the second moment 
 tight-binding model\cite{Ducastelle70}, incorporating ferro- and anti-ferromagnetic
 interaction terms which are needed for the FeCr system. For
 antiferromagnetic and paramagnetic materials, an explicit additional
 variable representing the spin is required.  In a mean-field 
 approximation this spin can be eliminated, and the potential becomes explicitly temperature dependent.
 For ferromagnetic interactions, this degree of freedom can be eliminated, 
 and the formalism
 reduces to the embedded atom method (EAM\cite{Daw84}), and we show the equivalence of
 existing EAM potentials to ``magnetic'' potentials.

\date{\today}

\section{Introduction}

Ferritic stainless steels have good radiation resistance are good
candidates for use in many nuclear applications.  A fundamental
understanding of their properties under radiation and ageing can be
obtained by molecular dynamics (MD) and Monte Carlo (MC) simulations.
The crucial ingredient for such studies is a model of the forces
between atoms: the interatomic potential.

While commercial stainless steels are typically multicomponent, the
major components are iron and chromium, which exhibit ferro- and
antiferromagnetism.  Early work neglected explicit treatment of
magnetism
\cite{Finnis84,Simonelli93,Ackland97,Mendelev03,Caro05,Olsson05,Muller07}
or was confined to a lattice\cite{Ackland06x,Ackland09}
since the Curie temperature $T_C$ of body-centred cubic (bcc) iron is
1043~K, this is a serious approximation.
Ab initio studies show that magnetic interactions between the atoms 
can have a major effect on defect energetics, e.g.
the vacancy formation energy changes by
up to 40\% with different magnetic states.
\cite{Korzhavyi99}.

A simple way to include magnetic effects involves defining 
environment-dependent
localised magnetic moments in total energy expressions\cite{Dudarev05,Ackland06,Ma08}. 
Here we extend this work and  outline a framework to develop a model treating magnetic
and non-magnetic Fe-Fe interactions in arbitrary configurations based
on the $2^{nd}$ moment approximation of tight-binding.

\section{Magnetism in the $2^{nd}$ moment approach}

\subsection{Band Energy}

We base our work on the second moment approximation to tight
binding\cite{Finnis84}, extended to multiple bands $s$, and two
$d-$bands of opposite spin.  In the two-band model the contribution to
the cohesive energy from the $d$-bands is proportional to the square
root of the bandwidth, which is approximated by a sum of pairwise
potentials \cite{Finnis84}.  By assuming local charge
neutrality\cite{AFV} and charge transfer between bands\cite{Ackland03,Olsson05}
multiple bands and arbitrary atomic numbers can be treated.

Consider a rectangular single spin band containing N states 
of full width $W$ centred on the
free atom eigenvalue (assumed degenerate).  
Electrons occupy the lower energy states, so relative to the free atom
the cohesive energy is given by:

 \begin{equation}  
U = \int_{-W/2}^{E_f}  \frac{5E}{W} dE =
 {Z} \left( \frac{Z}{5}-1 \right)\frac{W}{2} 
 \end{equation}

 where $Z$ is the occupation of the band and the Fermi energy is $E_f
= (Z/N-1/2)W$.  By defining a local density of states projected onto
an atom, this becomes an expression for energy per atom.  If we
consider a single band, $Z$ is constant for a given material, and the
cohesive energy varies parabolically with the number of $d$-electrons.
In the second moment approach, the bandwidth W can be written as the
square root of a sum of pairwise potentials, which following Finnis
and Sinclair\cite{Finnis84} we assume can be empirically fitted to data.

\begin{equation}
W_{b,i} = \sqrt{\sum_j \phi_{b}(r_{ij})} 
\end{equation}
where the fitting function $\phi_{b}(r_{ij}$ is a measure of the strength of
the bond. In transition metals, the cohesion arising from $d$-band formation is expected to
dominate, but the contribution from $s$-band formation may not be negligible \cite{Pettifor77}

To include magnetism, we assume two local $d$-band densities of states
(N=5) with opposite spins, labelled by arrow, and one $s$-band (N=2).
Writing down the second moment energy for three bands, one finds

\begin{equation} 
E^{band} =  U^\uparrow + U^\downarrow + U^s= W_d \left[ \frac{n_d^2+\mu^2}{20}- \frac{n_d}{2} \right]
+ W_s \left[ \frac{n_s^2-2n_s}{4}
 \right]
\label{dEn}
\end{equation}
where $n_d = Z^\uparrow + Z^\downarrow$ is the total number of
$d$-electrons, $n_s$ is the number of $s$ electrons.  The valence is
then $n_d+n_s$.  It is assumed that both d-bands have the same width
$W_d$, regardless of occupation.
The on-site magnetic moment 
(in units of Bohr magnetons) is then 
$\mu = Z^\uparrow - Z^\downarrow$. 

This band formation contribution to the magnetic energy is always
positive.  It can be shown that by a simple shift of variables it can  
be implicitly included in the many-body part of second-moment potentials.
This makes the ``Embedding function'' somewhat more complicated than the 
standard square root of Finnis-Sinclair type potentials, and helps
 explain the surprising success of previous EAM potentials developed
for iron \cite{Finnis84,Ackland97,Mendelev03}.

\subsection{Many-body repulsion}

In order to construct a potential usable in MD we need to add
 non-magnetic repulsive terms to the total energy expression. 
At equilibrium, s-band electrons contribute
 significantly to pressure and in particular bulk modulus
 \cite{Pettifor77}. Further, due to wavefunction overlap causing
 a shift of the centre of the
 d-band, d-electrons also contribute to repulsion small distances,
 although since d-orbitals are quite compact for transition metals
 this is expected to be small.  We therefore write for the on-site
 non-magnetic 2nd moment band energy:
\begin{eqnarray}
E^{nm} &=&  E_d^{rep}  +  E_s^{rep} + (2-n_s) E_{s\rightarrow d}
\end{eqnarray}

where $l_b$ is the orbital angular momentum of band $b$, $E_b^{rep}$
represents many-body repulsive interactions. $E_{s\rightarrow d}$
is the energy required to promote an $s$-electron to the $d$-level in
the free atom, a term which must be included if $s-d$ transfer is allowed. 

One possibility is that the many body repulsion term is approximated
by the kinetic energy of a free electron gas, where the local density may be measured by a sum of pair potentials, similar to the bandwidth.
\begin{equation}
E_{b,i}^{rep} = n_b^2 \rho_{b,i}^{5/3}; \hspace{1cm} \rho_{b,i} = \sum_j \phi^{rep}_{b}(r_{ij})
\end{equation}
Note that the function $\phi^{rep}_{b}$ will be different from  $\phi_{b}$ 
in equation 2.
It is expected that the $s$-electrons will give the major
contribution. The repulsion due to kinetic energy of $d$-electrons is
expected to be considerably smaller than the Pauli exclusion of
$d$-electrons with spins aligned in the same direction.

\subsection{Magnetic energy}

So far we have not considered interaction between spins, arising from
Pauli repulsion (exchange) and correlation.  This should have both
on-site and inter-site terms.  Expanding about the non-magnetised state,
 we choose a Landau-Ginzberg expansion 
for the on-site terms:
\begin{equation}
E^I_{i} =  - \frac{I_2}{4} \mu_i^2 + \frac{I_4}{8} \mu_i^4
\label{Emag}
\end{equation}
Where $I_{2}$ and $I_{4}$ are {\it on-site} spin-polarisation
 parameters following a Landau-Ginzberg expansion and  $E^{xc}$ is the
 {\it intersite} exchange-correlation energy. Partitioning into
pairwise contributions from repulsive Pauli exclusion between like-spin electrons ${\cal J}_{P}(r_{ij})$ 
and Heisenberg
 correlation  ${\cal J}_{c}(r_{ij})$ gives

\begin{eqnarray}
E^{xc}_i
  &=& \frac{1}{2}  \sum_j \left[{\cal J}_{P}(r_{ij}) \left(
  Z_i^{\uparrow} Z_j^{\uparrow}+Z_i^{\downarrow} Z_j^{\downarrow} \right)
+  {\cal J}_{c}(r_{ij}) {\bf  \mu_i \mu_j} \right] \nonumber \\ 
&=& 
\frac{1}{2} \sum_j \left[
{\cal J}_{P}(r_{ij}) \frac{(n_d^in_d^j)}{2}+ \left ( \frac{1}{2}{\cal J}_{P}(r_{ij}) 
+ {\cal J}_{c}(r_{ij}) \right ) {\bf \mu_i \mu_j} 
\right]
\label{XC}
\end{eqnarray}
where $n_d$ is the number of $d$-electrons in the solid.  These {\it
off-site} exchange-correlation interaction terms are essential for
modelling of Curie and Neel transitions as well as negative mixing
energies from frustration of anti-ferromagnetically aligned solute
atoms, such as Cr in Fe\cite{Olsson03,Wallenius04}.

The spins may be treated either dynamically (fictitious dynamics) or statically (Born-Oppenheimer approximation).  For the latter case
differentiating $E^{band}_i + E^{I}_{i}+ E^{xc}_i$, shows that the on-site magnetic
 moment $\mu_i$ is stationary in a perfect lattice (i.e. $|{\bf \mu_i}| =\mu$) when

\begin{equation}
\mu_i^2 =  \frac{1}{I_{4}} \left[I_{2} -\left( \frac{W_{d,i}}{5} + 
 \sum_j \left[ \frac{1}{2} {\cal J}_P(r_{ij}) + {\cal J}_c (r_{ij}) \right] \right) \right]
\label{momentsquare}
\end{equation}

In the free atom limit $r \rightarrow \infty$, Hund's rule
must be satisfied $\mu=\mu_{free}= 5-\left|5-n_d\right|$, and the
energy should be a minimum for this magnetic moment.  This gives us a
relation between the Landau-Ginzburg expansion coefficients and the
free atom properties:\[ I_2 = 8E_{free}/\mu_{free}^2 ; \hspace{1cm}
I_4 = 8E_{free}/\mu_{free}^4 \] where $E_{free}$ is the energy
difference between non-magnetic and magnetic free atom.
 $E_{free}$ and  $\mu_{free}$ are both numbers, easily measured from the free atom properties.

A further simplification can be made
using the Born-Oppenheimer approximation: the magnetisation is in the 
ground state for a given atomic configuration, and 
we may assume that $\partial E / \partial \mu = 0$.  This leads to

\begin{equation}
\mu_i =  \mu_{free} \sqrt{
  1 -\frac{\mu_{free}}{8E_{free}}\left( \frac{W_{d,i}}{5} +  \sum_j {\cal J}(r_{ij}) \right)
 }
\label{moment}
\end{equation}
where ${\cal J} = \frac{1}{2} {\cal J}_P + {\cal J}_c$; the two separate
physical effects are merged in a single pairwise interaction.  
Equation 10  shows that
as the canonical d-band broadens, lowering the band energy, so the
magnetic moment is reduced.  Thus the magnetic moment in the solid is
significantly lower than the free atom, the moment around a vacancy is
enhanced, and the moment around an interstitial defect is further
suppressed.

\subsection{Total Energy}
 
Collecting together the terms discussed above, the final form of the energy for our modified
second moment three band model with both on and off site magnetism is $E_{tot} = \sum_i E_i$, with:
\begin{eqnarray}
E_i & = & W_{d,i}\left [ \frac{n_d^2+\mu_{i}^2}{20} - \frac{n_d}{2} \right ]  + E_{d,i}^{rep} \nonumber \\
    & + & W_{s,i}\left [ \frac{n_s^2}{4} - \frac{n_s}{2} \right ] +  E_{s,i}^{rep}   + (2-n_s) E_{s\rightarrow d} \nonumber \\
    & + & \frac{1}{4}  \sum_j {\cal J}_P(r_{ij})n_d^i n_d^j +  \frac{1}{2} \sum_j {\cal J}(r_{ij}) \mu_i \mu_j \nonumber \\
    & -  & \frac{E^{\mu}_{\mathrm{free}}}{8}
\left[  \mu_i^2 -  \frac{ \mu_i^4}{2\mu_{free}^2} 
\right].
\label{TotalEnergy}
\end{eqnarray}
Once again ${\cal J} = \frac{1}{2} {\cal J}_P + {\cal J}_c$.  The model has 
numerical parameters  Z, $E_{s\rightarrow d}$, $E^{\mu}_{\mathrm{free}}$, 
requires pair potentials for  $W_{b,i}$,  $a_b^{rep}\phi_{b}(r_{ij})$,  ${\cal J}_P(r_{ij})$ and ${\cal J}(r_{ij})$.
It is interesting to note that the 
 s-band contributes to the bonding thanks to
the $s\rightarrow d$ transfer in the solid.  This increases the number of 
$d$-electrons, potentially increasing the magnetic moment for elements
with valence less than 7, and reducing it for atoms with valence above 7.

\subsection{Analytic expressions for charge transfer}

It is possible to treat $n_d$ and $n_s$ as variables, or to use the
Born-Oppenheimer Approximation and solve for $n_d$ and $n_s$ as
minimisers of the energy.  For the latter approach, we take the case of iron and  differentiate
the total energy (\ref{TotalEnergy}) with respect to $n_s$, under the
condition $n_d = 8 - n_s$. In a perfect lattice with zero net magnetic
moment 
$n_s$ is equal on all atoms $i$, we obtain :
\begin{equation}
n_{s} = \frac{W_{s} + \frac{3}{5} W_{d}+ 8 \sum_j {\cal J}_p(r_{ij})
+8 \rho_d^{5/3}+ 2E_{s\rightarrow d} } {W_{s}+ \frac{1}{5} W_{d} +
\sum_j {\cal J}_p(r_{ij}) + 4 \rho_s^{5/3} + 4 \rho_d^{5/3} }
\end{equation}

For magnetic states, the expression for $n_s$ becomes more complex:

\begin{equation}
n_{s} = 
\frac{
\frac
{W_s + \frac{3}{5} W_d + 8 \sum_j {\cal J}_{P,ij} - 4 \sum_j {\cal J}_{ij} + 8 \rho_d^{5/3} + 2E_{s\rightarrow d}}
{ \frac{1}{2}E_{\mathrm{free}}^{\mu} +
\left(6 (\sum_j {\cal J}_{ij})^2 + \frac{4}{5} \sum {\cal J}_{ij} - \frac{2}{25} W_d^2 \right)/E_{\mathrm{free}}^{\mu}}
+1}
{
\frac
{W_s + \frac{1}{5} W_d + 2 \sum_j {\cal J}_{P,ij} + 4 \sum_j {\cal J}_{ij} + 4 \rho_s^{5/3}  + 4 \rho_d^{5/3} }
{ \frac{1}{2}E_{\mathrm{free}}^{\mu} +
\left(6 (\sum_j {\cal J}_{ij})^2 + \frac{4}{5} \sum {\cal J}_{ij} - \frac{2}{25} W_d^2 \right)/E_{\mathrm{free}}^{\mu}}
-1} 
\end{equation}

This expression only applies in the homogeneous case, a mean-field
approximation when $\mu_i=\mu_j$.  Otherwise determining the $s-d$
transfer is non-local, and best treated by fictitious dynamics with $n_s$ and $\mu$
as
dynamical variables.  However, provided $\mu_i \approx \mu_j$ the error
is only second order.
To show this we can go to the low temperature 
limit where ${\bf \mu_i.\mu_j}= \pm \mu_i\mu_j$:
\begin{eqnarray} {\bf \mu_i.\mu_j} &=& \frac{(\mu_i+\mu_j)^2 - \mu_i^2 - \mu_j^2}{2} \nonumber \\
& = & \frac{(2\mu_i+\delta\mu_{ji})^2 - \mu_i^2 - \mu_j^2 }{4}  +  \frac{(2\mu_j+\delta\mu_{ij})^2 - \mu_i^2 - \mu_j^2 }{4} 
\end{eqnarray} 
where $\delta\mu_{ij}=\mu_i-\mu_j$
Now we can use the fact that the energy per atom is an arbitrary
quantity to assign terms in $\mu_j^2$ to $E_j$, making them local. Further
the cross term becomes 
\[ \mu_j.\delta\mu_{ij}+ \mu_i.\delta\mu_{ji} = (\delta\mu_{ij})^2 \]
i.e. second order in the difference in spin.  Thus the effect of 
inhomogeneities on $n_s$ is small.
At higher temperature the coupling term becomes even smaller.

With this redistribution of energy between atoms, we find
\begin{eqnarray}
 \sum_{ij} 
 {\cal J}(r_{ij}) ({\bf \mu_i.\mu_j})   
=
 \sum_{ij} 
 {\cal J}(r_{ij}) (\mu_i^2-\frac{1}{2}\delta\mu_{ij})   
\label{TotalEnergy2}\end{eqnarray}
To use the analytic expression for charge transfer, one has to ignore the
$\frac{1}{2}\delta\mu_{ij}$.

\section{Forces}
There are two approaches to applying the current model to MD.  For the
model with a full seven degrees of freedom per atom, one simply writes
a Lagrangian including $n_s$ and $\mu$ at each atom, defines fictitious
masses to each, and integrates the equations of motion.  However, it
is desirable to avoid explicitly including extra degrees of freedom.

The Hellmann-Feynman theorem tell us that to evaluate forces we do not have 
to consider partial derivative with respect to variational parameters.  
Here, in the Born-Oppenheimer approximation
$\mu_i$, $n_s$ and $n_d$ all fall into this category.  Thus the force is 

\begin{eqnarray} {\bf F_i} &=& -\frac{dE}{d{\bf r_i}} \nonumber \\
&=& -\frac{\partial E}{\partial{\bf r_i}} - \frac{\partial E}{\partial{n_s}}\frac{\partial n_s}{\partial{\bf r_i}}  - \frac{\partial E}{\partial{n_d}}\frac{\partial n_d}{\partial{\bf r_i}} 
 - \frac{\partial E}{\partial{\bf \mu_i}}\frac{\partial {\bf \mu_i}}{\partial{\bf r_i}}  \nonumber \\
&=& -\frac{\partial E}{\partial{\bf r_i}} \end{eqnarray}

Consequently, although the analytic solution for the energy in the
 Born-Oppenheimer approximation is cumbersome, it yields tremendous simplification of the forces.

\section{Equivalence to and comparison with two-band and  magnetic potentials}

If we take a filled $s$ band, $n_s$=2, neglect the intersite
interaction and take a pairwise potential rather than Fermi-gas
approach to the repulsive terms then we obtain the ``magnetic''
potential\cite{Dudarev05}.  Similarly, the proof that the two-band
magnetic model reduces in the Born-Oppenheimer
approximation to EAM\cite{Ackland06} can be trivially applied here under the 
same assumptions. 

Although previous EAM parameterisations for iron have not explicitly
included magnetism, since they are fitted to data which does include
magnetism, it should be possible to reverse engineer the atomistic magnetic
moment from any EAM potential.  We have shown here that 
it is simply a function of the d-band width, so we need only find the relevant relationship..

In figure 1 we show a comparison between atomic magnetic moments
calculated from first principles in iron\cite{Domain01,Fu04} and
bandwidth from an EAM iron potential \cite{Mendelev03}.  Also shown is the
equivalent figure for the Dudarev-Derlet magnetic
potential\cite{Dudarev05}.  The similarity between the two is
striking, but more interesting is the fact that the data does not fit
well to a single function of bandwidth.  It appears that a single
nearby interstitial has a much greater effect on suppressing the
magnetisation than it does in increasing the local bandwidth.  Closer
examination of the projected $d$-band density of states suggests that
its shape is changed by point defects, but not by compression. Since
the shape of the band is not properly captured
in second moment models.

\begin{figure}[ht]
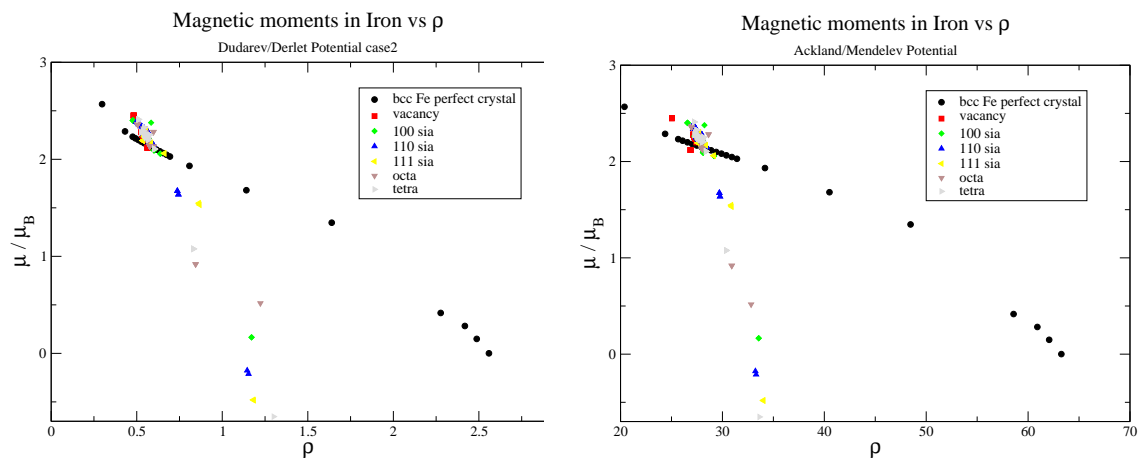

\protect{\includegraphics[width=0.45\columnwidth]{mm_vs_rho_Dudarev_case2.eps}}
\protect{\includegraphics[width=0.45\columnwidth]{mm_vs_rho.eps}}
\caption{Graphs of magnetic moment against $\rho$ ($W^2$) for Mendelev and Dudarev potentials. Symbols represent isotropic compression of bcc Fe, and sites close to point defect configurations including self interstitials in tetrahedral, octahedral and 
three dumbbell configurations }
\end{figure}

\section{Simplification: Elimination of the s-d electron transfer}

Computationally, the slowest part of the potential is the s-band.
This is because of the long range of the functions.  In practice, the
$s$-band occupation is relatively insensitive to the atomic positions,
and only weakly dependent on the density, and for simulation of solids
it is possible to fix the $s$-electron occupation without changing the
results.  This also has the significant advantage that the only
variable which needs to be optimised is the magnetisation: $n_d$ is
constant.  On account of its partial occupancy, the $s$-band still
contributes significantly to the cohesive energy, an effect ignored by
previous models, but its long range means it has little effect on
structural properties.  Fixing $n_s$ has the important practical
advantage that the $\mu_i$ can be determined algebraically.  This is
only likely to be problematic when considering properties involving
the free atom, e.g. cohesive energy or sputtering.

A still further simplification eliminates the s-band terms altogether.
In a practical molecular dynamics simulations, for a reasonably
homogeneous system, the free-electron s-band width can be written a
simple non-local function of density, taking the same value for all
atoms.  In a constant volume simulation, this will be the same for all
time, and simply add a constant term to the cohesive energy and bulk
modulus. It suffers from the problem that the bulk modulus associated
with volume change is different from that calculated from the limit of
long-wavelength phonons (and therefore one needs to decide which to fit), but in many practical applications this 
may be unimportant.

\section{Simplification: Decoupling the magnitude and coupling of $\mu$}

The time and energy scales for rotation of the spins are much smaller
than for changes in spin magnitude or atomic position.  This means we
can obtain some simplification by integrating out the rotational
degrees of freedom.

The intersite spin term, ${\cal J}(r_{ij}) {\bf \mu_i.\mu_j} $, couples
spins on adjacent sites.  This makes the task of evaluating ground
state for the full set of spins $\{ \mu_I \}$ nonlocal and slow: in
practice it is more sensible to evaluate the spin degrees of freedom
explicitly and dynamically\cite{Ma08}.  It is
possible, however, to greatly simplify the potential by making a
pairwise approximation for the spins.  Here we store only the
magnitude of the spins on each site, and replace the explicit dot
product with its ensemble average, treating each ``bond''
independently.

\begin{eqnarray}    
 {\bf \mu_i \cdot \mu_j}  & \rightarrow& 
\frac {
\int{
 \mu_i\mu_j \cos \theta  e^{ {\cal J}  \mu_i\mu_j\cos\theta/kT} 
\sin \theta d\theta d\phi }
}{
\int 
e^{ {\cal J}  \mu_i \mu_j \cos \theta /kT } 
\sin \theta d\theta d\phi
}\\
&=&
\coth \left (\frac{J\mu_i\mu_j}{kT} \right )-\frac{kT}{J\mu_i\mu_j}
\end{eqnarray}

The advantage of this approach is twofold: it reduces the degrees of
freedom from 6N to 4N, and by treating the fastest processes
implicitly it allows one to use a larger timestep.  A curious feature
is that the potential becomes explicitly temperature dependent.
Unlike simple magnetic potentials\cite{Dudarev05,Ackland06,Dudarev08} it still
allows one to distinguish local ferromagnetic and antiferromagnetic
behaviour, but not antiferromagnetic frustration.

\section{Conclusions}

We have presented a formalism within which magnetism can be treated
within the second moment approximation to tight binding, a Heisenburg
model for spin-coupling and and a Landau-Ginzberg approximation for
onsite spins.  The model is mathematically complicated, at worst
requiring four additional variables ($n_s$ and a vector spin) per
atom.  We have also shown, however, that these variables can be
eliminated with well controlled approximations.  Computationally, the
model is efficient, requiring only the evaluation of short range pair
potentials for forces or energies.

Parameterisation of the model is a significant undertaking. In
principle the various parameters and pair potentials must be fitted
for each element on pair of elements.  It may prove that similar
functions apply across the whole group\cite{Hepburn09}, and the
division of the energy into magnetic and non-magnetic parts allows
data calculated using either method from first principles to be used.
The combination of a mean-field approach to the inter-site magnetism,
the Born-Oppenheimer approximation for the spin degree of freedom and
the Hellmann-Feynman theorem for forces reduces the model to a
temperature-dependent EAM, albeit with more than one embedding term.

The model automatically reproduces a number of features of transition
metal binding: band formation, magnetism, $s-d$ charge transfer,
suppression of magnetism in forming a solid, further suppression of
magnetism in overcoordinated defects and increased cohesion for
elements in the centre of the $d$-series.  Since each term has a
well-defined physical interpretation within the local density of
states in tight-binding band theory, results from the model provide 
physical intuition as well as numerical predictions.

The authors acknowledge support from the EU under the GETMAT project.

\end{document}